# The Stength of Weak cooperation:
# A Case Study on Flickr

Christophe Prieur[(a+b)], Dominique Cardon[(b)],
Jean-Samuel Beuscart[(b)], Nicolas Pissard[(b)],
Pascal Pons[(b)]

[(a)] LIAFA, Université Paris-Diderot
Case 7014
75 205 Paris Cedex 13

prieur@liafa.jussieu.fr

[(b)] SENSE, Orange Labs
38 rue du général Leclerc
92 131 Issy Moulineaux
00 33 1 45 29 57 74

domi.cardon@orange-ftgroup.com

## ABSTRACT
Web 2.0 works with the principle of weak cooperation, where a huge amount of individual contributions build solid and structured sources of data. In this paper, we detail the main properties of this weak cooperation by illustrating them on the photo publication website Flickr, showing the variety of uses producing a rich content and the various procedures devised by Flickr users themselves to select quality. We underlined the interaction between small and heavy users as a specific form of collective production in large social networks communities. We also give the main statistics on the (5M-users, 150M-photos) data basis we worked on for this study, collected from Flickr website using the public API.

## 1. INTRODUCTION
Without trying (once again) to define what lies (and what does not) behind the label "Web 2.0", one can at least deal with the articulation of individual self-production practices and cooperation between Internet users, resulting in the collective construction, on the WWW, of big, structured sources of information made of a huge amount of individual contributions.

The development of the 'good-old Web' had always been driven by a community ideal, and it had been built up mainly through organized cooperation between voluntary participants. In this context, the cooperation between members has often been described as strong: mutual socialization and defined roles give members a feeling of belonging to the community and a joint, shared aim [1]. The successful growth of Web 2.0 services (driven by Wikipedia, blogs, Flickr, etc.) has led to the definition of a much weaker cooperation between Internet users, detailed in [1].

As a result of the spread of self-production tools (image, video, blog platforms, wiki, etc.), Web 2.0 services enable cooperation between Internet users as a side effect of their individual publication activities. The 'strength of weak cooperation'[1] lies in the fact that it is not necessary for individuals to have a cooperative plan of action or an altruistic concern beforehand. They discover cooperative opportunities simply by making their individual productions public. Public space is seen as an opportunity for one's visibility, leading to relation making and eventually actual cooperation with different levels of involvement. And this cooperation can work in a very large scale precisely because it is non-demanding. This weak cooperation in a numeric space also allows cooperation between small and heavy users which could be problematic in real life.

As a website for photo publication providing tools that enable coordination, Flickr is often showed as a typical example of the Web 2.0 [4]. The aim of this paper is to detail the concept of weak cooperation on this example, showing the great variety of uses, from plain stockpiling of photos to complex combinations of all the functionalities, and how these functionalities serve both individualistic purposes such as building one's notoriety and altruist ones since they lead to a highly structured base of photos with many user-generated procedures to select quality from quantity.

We first describe in Section 2 the functionalities of the website and the database we used for our study, giving basic figures on the uses of the website. Section 3 deals with individual aspects of these uses such as the variety of individual practices and the necessity of 'playing the game' to get acknowledged. This last point leads to Section 4 where collective issues are addressed, studying the user-created groups, mixing a both thematic and social functionality whose role in the weak cooperation is crucial since it enables users to invent their own procedures of selection.

## 2. FLICKR, SYMBOL OF WEB 2.0
Although Flickr is among the original 'officially' Web 2.0 websites[2], its founders had not anticipated that it would become a photo publication tool. Stewart Butterfield and Caterina Fake (see [4]) initially intended it as a multi-player game, then as a platform with chatrooms where people would share objects materialized as pictures. But uploading of personal pictures took more and more importance in the service launched by Ludicorp in February 2004. The functionalities evolved to suppress chatrooms and provide personal pages to users. After a few months of growing success, Flickr was acquired by Yahoo! for reportedly $30 million.

---

[1] The expression is of course coined in reference to Granovetter's S*trength of Weak Ties* [3].

[2] in the exemplified definition seminally given by Tim O'Reilly [3].

The ability of the creators of Flickr to follow the actual uses of their service was a key to its original positioning and thus to its success. It came at just the right time, combining the boom of the sales of digital cameras, the growth of social networks services and the success of blogs, for which Flickr soon provided posting tools.

Some studies have already been done on Flickr. The history of the site and its emblematic importance in the web 2.0 paradigm has been introduced by Cox [4] and Van House [7], while Marlow *et al.* [8] present Flickr as an example of Folksonomy systems. But few studies are based on extractions of Flickr database. To our knowledge, the only large statistical analysis of Flick data has been done by Kumar *et al.* [9] at Yahoo. They present a series of measurements of the evolution of the different components of Flickr's relational structure. In this seminal work they have demonstrated that Flickr (and Yahoo! 360) is composed of a growing giant connected component (59.7% of users at the end of the studied period) that represents the large group of people who are connected to each other through paths in the social network. Beside this giant component, they describe a middle region of less connected users, and then isolated singletons. While this structure is characteristic of large networks, they show that the proportion of the middle region is constant over the time, taking 1/3 of the users. In another context, Lerman and Jones [10] extracted small samples of data from Flickr in order to show the role of contacts for browsing on the site. The most important part of the studies on Flickr deal with the analysis of the evolution of photographic practices [7]. In an examination of digital photographers' "photowork activities" [11], Miller and Edwards [12] have showed that for some people, Flickr supports a different set of photography practices, socializing styles and perspective on privacy than traditional photo amateurs. Our study comforts this idea that in transforming amateur practices in a public activity, Flickr has proposed a new paradigm for amateurs in which reputation and visibility can be built by the intensity of the communicative involvement with Flickr functionalities. Since Chalfen [13] early book about the "Kodak Culture" of amateur photography, the rise of Internet-based photo-sharing has strongly affected domestic practices of photography. In Kodak Culture, a small group of persons (friends and family) share oral stories *around* images with others. In the new culture of image – called "Snaprs" (a reference to the missing "e" in Flickr) by the authors – photos are used to tell stories *with* images, rather than *about* images as with the home mode [see also 14, 20]. In this new context, photo is not a story shared with closed relatives, but a large-scaled conversation shared with people that participants don't know in real life. Our study shows that Kodak and Snaprs cultures coexist on Flickr platform, but that Snapr users lead the community.

## 2.1 Main functionalities

Photos are the center of Flickr's activity. Users can index them with **tags** (freely chosen keywords), post them to thematic user-created **groups**, and put **comments** to them. Only the owner of a photo can post it to a group, while any user can tag and comment other users' photos. Users can also mark as **favorites** other users's photos.

Users have to register to a **group** to be able to post photos to it and users can mark other users as **contacts**.

Basic membership is free but has some limitations with respect to a paying so-called "pro" account (only the last 200 uploaded photos of the user are displayed, the user can only create three sets, and the per-month upload bandwidth limit is lower[3].)

## 2.2 Harvesting the data

During Summer 2006, we have used the Flickr public API[4] to extract all *public* data concerning the five functionalities listed above (tags, groups, comments, favorites, contacts). For users, only the identifiers have been stored (no personal information) and for photos, only identifiers and titles (of course not the photo itself).

The extraction was done (in Java) by iterating on each user id $u$, to get all contacts and (public) groups of $u$, and by iterating on each (public) photo $p$ of $u$, to get all comments, tags and favorites of $p$. Another iteration was then done on each group $g$ to get the list of photos posted in $g$.

## 2.3 Basic figures

By definition, private photos are… private, thus unreachable by the API. However we can give an upper bound for their quantity since the ids of Flickr photos are numbered by upload order[5]. For instance, we have in our photo base the ids 222851183 and 222851185 but not the 222851184. The latter is thus either private or has been deleted. By this mean, one can claim that private photos are not more than 33%[6] (since the ids that we have in our base cover 67% of the range). In the rest of the paper, only public photos will be considered unless specifically mentioned.

Figure 1 shows the distribution of photos, which is of course highly heterogeneous (although technically not in power law), 20% of the users owning more than 82% of the photos. One counts 156 840 996 (public) photos for 4 788 438 users, which makes an average of 33 for all users or 87 for users having at least one photo. "Pro" accounts have naturally much more photos: they own 59.5% of photos while they represent 3.7% of the users.

---

[3] Since our data extraction, these rules have changed and pro accounts don't have upload limit any more.

[4] http://www.flickr.com/services/api/

[5] Let us just mention for the anecdote the first public photo, numbered 74, http://www.flickr.com/photos/bees/74/, uploaded on December 15, 2003 and named big_test.

[6] In an interview given in April 2005, Stewart Butterfield even gave an 82% for public photos (see [12]). Note that our 67% is rather constant in time (actually it goes between 65% and 70%), which does not contradict the 82% since we don't have a way to know the amount of uploaded-then-deleted photos.

| functionalities | | total | per user (having at least 1) | | | % of users having 0 | | |
|---|---|---|---|---|---|---|---|---|
| | | | all | non-pro | pro | all | non-pro | pro |
| photos | | 156 840 996 | 87 | 39 | 562 | 62 | 65 | 6 |
| contacts | of a user | 14 926 127 | 9 | 6 | 40 | 65 | 67 | 20 |
| | incoming | - | 9 | 6 | 41 | 65 | 66 | 16 |
| comments | given | 46 646 865 | 76 | 26 | 254 | 87 | 90 | 25 |
| | received | - | 61 | 24 | 271 | 84 | 86 | 35 |
| favorites | given | 17 883 026 | 56 | 27 | 145 | 93 | 95 | 56 |
| | received | - | 52 | 15 | 131 | 93 | 95 | 39 |
| groups | | 72 875 | 15 | 8 | 37 | 92 | 94 | 51 |

**Table 1. Distribution of Flickr Functionalities**

One reads this table as follows: the average number of photos per user having at least 1 photo is 87 among all users, 39 among non-pro, 562 among pro users. Users with no photo make 62%, 65% and 6% respectively among all users, non-pro users and pro users.

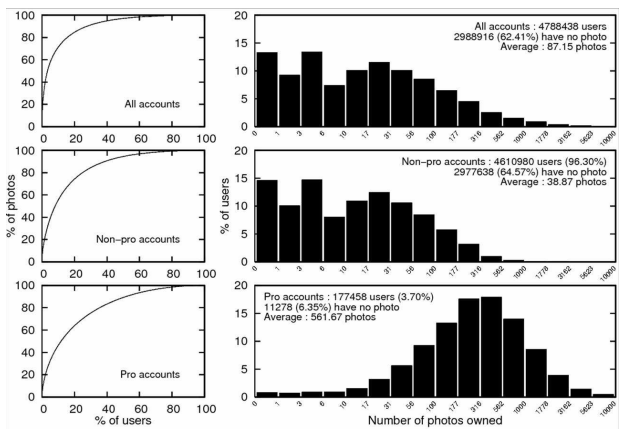

**Figure 1. Public photos per user**

Table 1 above sums up the average use of each functionality. The first obvious thing to remark is the big difference between pro and non-pro users, which is not only a consequence of the limitation in the number of photos, since it can also be observed on the average number of contacts (6 vs. 40). What can also be noticed is the different amount of uses of the various functionalities, even among pro users (only less than half of pro users use groups or favorites). Before studying in detail Section 3 this diversity, let us focus on the most active users.

### 2.4 Top sample base

As we have just seen, the activity of Flickr users is very heterogeneous in intensity. In order to study the particularities of the social uses of the site, we have extracted a sample base for some of the measures presented in the next section[7]. This base is made of the 50 000 more intensive users, where the intensity of a user was measured by taking the sum of the normalized ranks of a user on each of the functionalities. In this base, the average number of posted photos is 915 (with a maximum of 75 737), of contacts 181, of favorites 270 (received 307), of posted comments 775 (received 751).

## 3. BUILDING ONE'S REPUTATION

### 3.1 Various public uses

The originality of Flickr was to mix photo storage facilities with social activity. Figure 2 shows the repartition of the usage of the functionalities among all registered users.

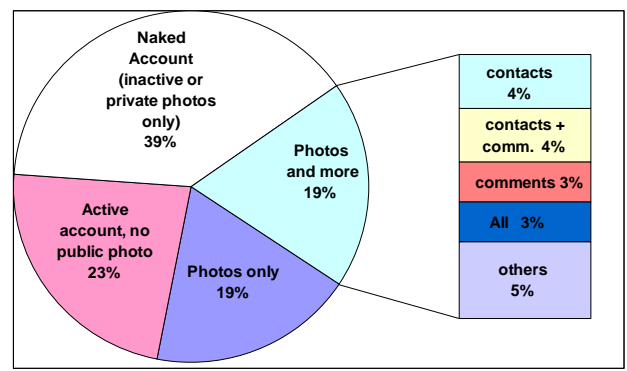

**Figure 2. Distribution of functionalities among all registered user**

First, 39% of the registered users seem to be totally inactive. They don't use any functionalities of the site and they haven't uploaded public photos. Second, 23% of users haven't uploaded public photos but have used communication functionalities of the site. We could hypothesize that a small part of participants of those two categories of users have uploaded private photos that we couldn't catch in our data. Nevertheless, their participation in general Flickr activities remain very small even if they represent 62% of all registered users. The strong heterogeneous distribution of the intensity of participation is a common law of all web (2.0 or not) platforms. In the following sections, we will only discuss on the remaining 38% of users. We could distinguish two groups of users: 19% of them upload public photos without using communication functionalities and 19% have both uploaded public photos and used various communicative functionalities such as contacts, comments or group participation.

---

[7] We always mention which measures are done on the whole base and which on this top sample base.

This opposition strengthens the main difference in Flickr practices between people using Flickr in order to store their own private or public pictures and those who use photographs as a way to communicate with others. As it has been described in many other online platforms such as Wikipedia [16], Blogs [17] or YouTube [18], a very small minority of users produce a large amount of the content but also organize this content through their activities: creating or animating group, tagging pictures, organizing contests, defining reputation of others, etc.

As soon as we concentrate our observation on these users, we can observe two significant kinds of social networking practices. Some are more interested in social contacts, others by socializing content. Those results can be observed with the correlation matrix of the uses of Flickr functionalities (Table 3 on the next page, we discuss this more in detail below): social practices such as incoming and outgoing contacts are strongly correlated with each other, but not with the number of uploaded pictures. On the contrary, sharing comments or favorites are closely linked together and also strongly associated with the number of photos.

|  | Component | | |
|---|---|---|---|
|  | 1 | 2 | 3 |
| **Nb photos** | -0.56 | -0.238 | 0.615 |
| **nb contacts out** | 0.325 | 0.833 | 0.058 |
| **nb groups** | 0.196 | 0.058 | -0.771 |
| **nb contacts in** | 0.648 | 0.662 | 0.191 |
| **nb favorites out** | 0.529 | -0.211 | -0.207 |
| **nb favorites in** | 0.808 | -0.058 | 0.097 |
| **nb comments out** | 0.894 | -0.277 | 0.085 |
| **nb comments in** | 0.720 | -0.443 | -0.003 |

**Table 2. Three dimensional PCA: three type of uses[8] (top sample base)**

To be more precise, Table 2 summarizes the result, on the top sample base, of a principal component analysis in three dimensions showing three types of uses, the first one opposing the number of photos to the rest of the functionalities (*social media use*), the second one opposing the functionalities attached to photos to the functionalities attached to the user (*MySpace-like*) and the third where most of the activity is concentrated on uploading photos (*photo stockpiling*). A synthetic projection on the first two components is given on Figure 3. In the last type of use (photo storage), people upload photos but have no communication practices with other users. In this context, Flickr appears only as a personal repository. We could hypothesize that most of them belong to the "Kodak culture" [13] that can be characterized by holiday and family pictures. The second one is a kind of MySpace-like use of Flickr. People upload a small number of pictures but have an intense use of communication functions. They use Flickr as a social network site in order to find new friends, sometimes with no clear links with (public) photographic activities. The first type of use is the conversational use of photography which characterizes the

---

[8] The three axis of the analysis explain 68% of the variance, which means that it is rather reliable.

"Snaprs Culture" [13]. In this context, people share contents, comments and social relations. This variety of uses shows the flexibility of the platform. But it also demonstrates that a minority of active users can lead the whole community (see also [9]).

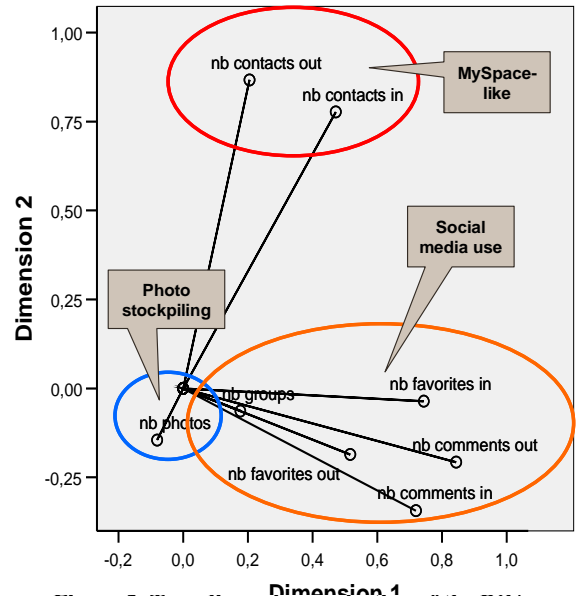

**Figure 3. Two-dimension projection of the PCA**

### 3.2 Reciprocity

The core principle of "social media" is that the individual practices just described are driven by the recognition users give to each other. It is no surprise that a high part (64%) of the contacts are reciprocated. This reciprocity is 32% for comments between users (*i.e.* the fact that a user *u* has commented at least one photo of user *v*), which is still very high since contrarily to contacts, returning a comment requires more than clicking on a link: you have to go to the user's page, chose a photo and… find something to say. Table 3 shows the correlations between the different functionalities. The highest correlation value (0.87) is precisely for received comments *vs.* posted comments, which means that people posting many comments also receive many comments.

| correlations | nb photos | nb groups | nb outgoing contacts | nb incoming contacts | nb faves granted | nb faves received | nb comments posted | nb comments received |
|---|---|---|---|---|---|---|---|---|
| **nb photos** | 1,00 | | | | | | | |
| **nb groups** | 0,24 | 1,00 | | | | | | |
| **outgoing contacts** | 0,13 | 0,45 | 1,00 | | | | | |
| **incoming contacts** | 0,17 | 0,51 | 0,76 | 1,00 | | | | |
| **nb faves granted** | 0,17 | 0,46 | 0,30 | 0,39 | 1,00 | | | |
| **nb faves received** | 0,16 | 0,42 | 0,28 | 0,61 | 0,47 | 1,00 | | |
| **nb com. posted** | 0,20 | 0,52 | 0,36 | 0,60 | 0,53 | 0,78 | 1,00 | |
| **nb com. received** | 0,17 | 0,49 | 0,29 | 0,47 | 0,53 | 0,55 | 0,87 | 1,00 |

**Table 3. Correlations between functionalities per user**

Of course this doesn't mean any general rule: more than 2 300 users have posted at least 100 comments without having received only one, whereas only 317 users have *received* at least 100 comments without having posted any. Posting is always easier than receiving… The difficulty is even greater for favorites since this functionality is by definition a matter of taste: only 13% of favorites between users (user $u$ has marked at least one photo of user $v$ as favorite) are reciprocated and the correlation between favorites given and received is very low (0.47). However, an interesting clue for favorites, as will be detailed now, is the high correlation (0.78) between favorites received and *comments* posted, suggesting that if you don't necessarily get "faved" by commenting other people's photos, at least you will be much more likely to. Note that this is confirmed by the fact that for the users in our top sample base, the average number of favorites received is even (slightly) greater than the one of favorites given.

### 3.3 Flickr's star system

Since the Flickr platform provides visible signs of recognition (views, faves, comments), it generates a sub-population of star photographers, characterized by very good audience figures (up to 1 million views, 100 comments per photo) often combined with other forms of recognition.

To have an insight on how Flickr stars are made, we tested a simple regression model on our top sample base. The dependant variable is the number of favorites received. We explain it with the variables of activity on Flickr: photos posted, comments made, favorites granted, groups membership, contacts made. The regression analysis suggests ($R^2=0.51$) that the best way to obtain gratifications is to post a lot of comments, then comes giving favorites and participating in groups. Social activity is a necessary condition to reputation: one of the prominent Flickr stars is also the top commentator on the site (51 400 comments posted in 18 month of activity).

So making oneself visible by posting a lot of comments, and a lot of photos into groups, is one of the keys to success on Flickr. Fame and recognition can also be earned or maintained in the editorial ecosystem developed around Flickr: a large variety of blogs, groups, user-made algorithms, work at extracting the *crème de la crème* of Flickr, providing selection of photos, interviews with Flickr artists, thematic selections, etc. In return, stardom on Flickr leads the elected users to intensify their practice. For some users, Flickr fame is converted into real-life recognition and benefits, like publications in magazines, exhibition, and professional opportunities.

*"I can honestly say I never, ever expected, when I first started using flickr to simply keep my drawings somewhere online to easily be able to show them to friends, that I would end up becoming one of the most popular people on flickr […] I'm amazed and quite touched at how many people regularly visit my photostream, it's gotten 875 000 views in less than a year, and that's just an absurd number to me. I mean, Iceland, where I live, has only 300 000 inhabitants! So this has been a very cool experience for me, I've started getting attention here in Iceland as well, which makes this all seem more real somehow. I'm very optimistic about the future. I am currently studying visual arts, preparing an exhibition, and I got my first paying shoot"* (Rebekka, http://flickr.com/people/rebba/).

Even though Rebekka, quoted above, may have created her Flickr account with the idea of being a stockpiling-type user, her publication activity was for her an opportunity of interaction and as she started to "play the game" of the social media, she became so involved that she is now part of the lead users who operate this weak cooperation.

## 4. GROUPS, A COORDINATION TOOL

The contact functionality is one-to-one. Functionalities attached to photos (comments, tags, favorites) are essentially one-to-many, even though some photos' comments may be the occasion for discussion between commentators. The place for many-to-many interactions is groups. In groups users can interact independently of a photo or a photograph, have discussions or make decisions on photos, photographers, groups or even Flickr. The fact that only 8% of all Flickr users (49% of pro users) are in groups is again a mark of the weak cooperation, where an active minority operates the structuring of the whole community.

Figure 4 shows the distribution of the number of members and of photos among the 72 875 groups. Technically, a group is made of a *pool* of photos posted by users who have previously joined the group, and of a *discussion forum* where messages may include small versions of photos (taken in the pool or elsewhere on Flickr). What makes groups an important tool is their flexibility: any user can create a group, decide the rules governing the posting of photos to the pool and of messages to the forum, and name administrators who will be responsible for the application of these rules[9]. There is thus a great diversity in the types of groups, in their content as much as in their rules and activity.

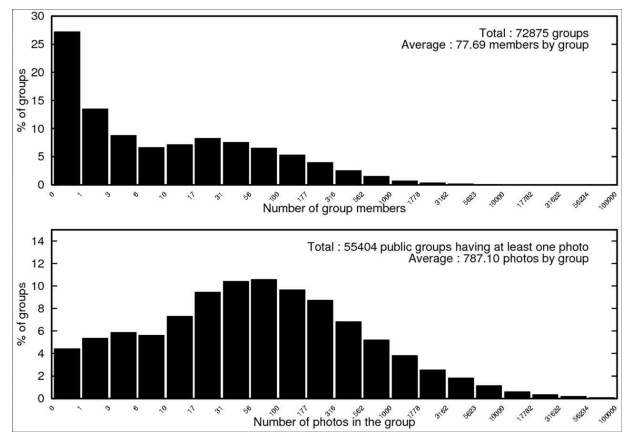

**Figure 4. Distribution of Flickr groups**

### 4.1 Thematic *and* social tool

Among the site's functionalities, tags, contacts and groups are the three giving direct access to photos. The first two have very distinct functions: tags are essentially used for indexing — a photo with the tag *cat* will appear in global searches made on this tag. As for contacts, they are the core material of the *social media* — Flickr shows you the recent photos of your contacts with the idea that people don't only want to see photos of *something* but also *someone's* photos [10]. Now groups draw on both aspects: they gather not only photos on one topic but also people, who contribute (or not) to give a social identity to the group by their activity.

---

[9] Unlike other members, the administrators of a group have the technical possibility to remove photos, forum posts or even members from the group.

This wide range of group types partly explains their very high thematic redundancy (over 300 groups about just cats). The simplest are defined virtually around a tag (cat, Paris, etc.) with no publication restrictions or specific activity. Their interest lies mainly in increasing the chances of photos being seen. Conversely, in some groups photos are a pretext for abundant discussions on the forum or for playing games with them. In the group **Flick-O-System: ? degrees of separation**, each discussion thread is a game with photos [10] (not necessarily taken from the group's pool): a thread where each photo shares a small detail with the previous one, another thread with characters looking alternately right and left (see picture on the… right). This sociability within a group sometimes extends to physical meetings, like in the group **flickr@paris**, *"Where the parisian and tourist flickrites meet, party, and get some pictures done together... Places change often, dates too, so keep an eye on the topics announcing events"*.

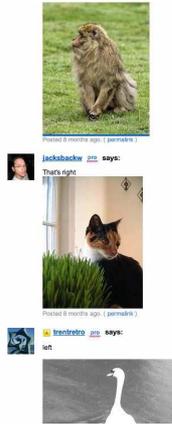

Of course many groups are somewhere between mostly thematic and mostly social, since making social activity from a thematic goal is easy. To take an example, the group **The Moon [*current* photos only]** is so specifically thematic (*"Please do not post any pictures of the Moon that are older than three days old in this group, pictures older than that will be deleted. Pictures that don't contain the Moon will also be deleted"*) that its administration itself becomes social activity, whose tracks can be seen in the forum.

### 4.2 An analytic scheme

In order to draw a map of the groups following the two aspects just described, namely tags and contacts as respectively thematic and social indicators[11], let us present briefly two measures of these[12].

Given a group $g$, we will call the *thematic graph* (resp. *social graph*) of $g$ the graph whose *vertices* (*i.e.* nodes) are the members of $g$ having posted at least one photo with at least one tag, and where an (undirected) *edge* (*i.e.* link) between users $u$ and $v$ denotes the fact that they have at least one tag in common (resp. one is a contact of the other). Thematic edges will be weighted using a function $w$ defined as follows.

- Given a tag $t$ and a user $u$, $n_t$ and $n_t(u)$ denote respectively the number of all Flickr photos and the number of photos of user $u$, both having tag $t$ (including photos outside studied groups). The maximal value of $n_t$ is denoted by $n_{max}$.

- The *rarity coefficient* $\rho_t$ of a tag $t$ is defined by $log(1+n_{max}/n_t)$. This coefficient ranges from 1 for the most used tag *beach* to approximately 10 for the rarest ones.

- The *tag weight* $w_{u,t}$ of tag $t$ on user $u$ is defined by 0 if $n_t(u)=0$, by $1+log\ n_t(u)$ otherwise. The idea of the *log* is of course to reduce the impact of users posting thousands of photos about the same topic (their wedding, baby, cat, holiday...)

- Finally the *edge weight* between users $u$ and $v$ is: $w_{u,v} = w_{v,u} = \Sigma_t\ (\ \rho_t\ \times\ min(w_{u,t},\ w_{v,t})$, which is meant to tell whether u and v share many tags, taking into account the rarity of these tags: the rarer are the tags, the closer the users are to each other.

Let us now recall that a *Lorentz curve* graphically shows a cumulative distribution function (the leftmost curves on Figure 1 in Section 2 are Lorentz curves) and that the *Gini coefficient* of a distribution is the area between the Lorentz curve and the diagonal (which is the Lorentz curve of the uniform distribution). This coefficient is a measure of the heterogeneity of the distribution: in the case of the number of photos owned by members (Figure 1), the Lorentz curve for pro users is closer to the diagonal than the one for all users, thus the Gini coefficient (thus the heterogeneity of the distribution) is lower.

We will now label a group by its *social density*, defined as the *density* of its social graph (*i.e.* the ratio of existing edges among all possible edges given the number of vertices) and its *tag dispersion*, defined as the Gini coefficient of the distribution of edge weights in its thematic graph. Figure 5 shows the results for a sample of the 450 groups having between 433 and 500 members (in our database, thus at the time of the crawl).

What is interesting is to look at the groups lying away from the upper-left cloud of mainstream groups with low social density and high tag dispersion. The most thematic ones, whose position is in the lower part of the chart, are listed on the left-hand side of the chart. Three-quarters of these group are in two categories: geographical, especially cities (Buenos Aires, Tel Aviv, Taipei etc.) and technical groups (K750i, XPRO, Fuji etc.), whose social densities range from very low values (Vienna, Stockholm for cities, K750i, expired films for technical) to quite high ones (Tel Aviv, Buenos Aires and toycamera, XPRO). In the case of cities, the social density may distinguish between tourism groups (where people just post photos of their travels without having much contact with others) and everyday-life groups, as suggested by the name of the group **Tel Aviv Stories**.

---

[10] One can see there the spirit of the initially intended Flickr as recalled earlier in this paper.

[11] Of course these criteria are used as a proxy. In many circumstances, the contact functionality is used as a bookmark to a user's photos, which may thus also indicate a thematic relation. As for tags, many are used precisely by groups as an identity (thus social) mark (*deleteme1*, *top-f25*…).

[12] See [13] for more details.

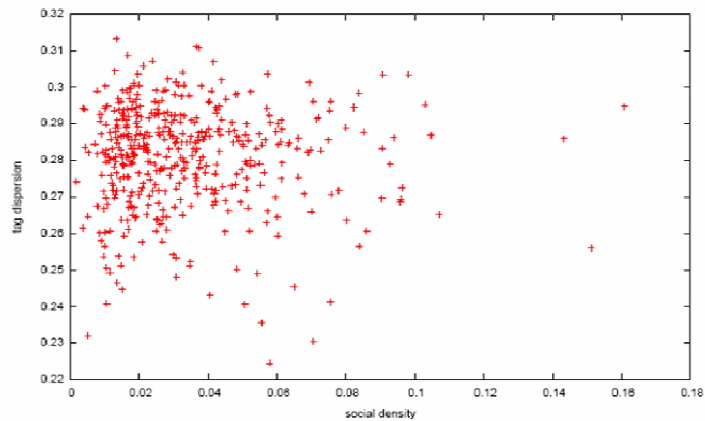

Figure 5. Social and thematic indicators for 450 groups
soc is for social density, thm for tag dispersion

As for groups with high social density, listed on the right-hand side of the chart, let us discuss on the first three easily distinguishable on the far right on the chart. The group **Paralelas/Parallels** is intended for photos with… parallel lines (wires, skyscrapers etc.), which could mean any kind of photos (the tag dispersion is high). But as suggested by the title in Portuguese, many members are from Brazil. This is an example of a social group whose social activity comes from a geographical proximity of its members (as was the case for Tel Aviv Stories). The group **FLICKRGAYS** is one of the (quite few) examples of both thematic and social groups[13] and may have some relevance in terms of social cohesion. Finally, **Fifty Faves** is for photos having been marked as favorites by at least fifty users. Of course not thematic, this group is for very experienced Flickr users, who know each other and have discussions about their productions. Along with many similar groups (**top-f50**, **GreatPixGallery 100faves+**, **100 club** etc.), it can also be seen as a popularity enhancer and one of the numerous groups whose function is to select quality.

### 4.3  What does quality mean?

The editorial function is a response to the need for quality in a context of decentralized self-production. Many groups are created with this purpose, with again various ways to achieve it. Some highly prestigious groups set themselves up as very selective, "*heavily curated*" galleries, to quote the warning given in the description of the group **Hardcore Street Photography**, which refers to professional photo agencies as models and rejects photos without explanations (*"we don't have a quantified set of rules. It's just a feeling that we have"*).

There is also a large family of voting groups, working on the following principle: each time someone posts a photo, they must rate or comment on one or more photos of the group[14]. The administrators just delete the photos of members who do not play the game. Besides enabling an automatic feedback for one's photos, some of these groups also have a 'select' double, intended for photos having successfully gone through the voting process. As an example, in the group **DeleteMe!**, members are invited to tag photos with either *deleteme* or *saveme*. After ten *deleteme*, a photo is deleted from the pool. After ten *saveme*, it is invited in the group **THE SAFE**[15], where it is voted on in a weekly thread of the forum, along with all photos 'saved' during the same week.

Even though this example is particular among the family of voting groups, since it is essentially devised as a game (*"On flickr we are all nice and sweet... always with a tender word for a flickrbuddy... [In the DeleteMe! group,] time to be nasty, mean, selfish and arrogant, time to dare to say what we think... and nobody can complain because that's the rules members accept. […] So just dare to put some of your photos to see how we appreciate them and how quick we will remove it from the group"*), it still illustrates the kinds of sophistication that can be reached by procedures devised by Flickr users to select quality, as was studied in [12].

What is most remarkable is that all these procedures might be seen as redundant with a built-in functionality of the service, namely the *interestingness*, a kind of pagerank for Flickr photos, taking into account elements such as the popularity of who has viewed them, marked them as favorites etc. This redundancy shows that there is not one unique measure of "interestingness" or of quality and people appear to want some control on what kind of aesthetics they want.

### 5.  CONCLUSION AND FUTURE WORK

In this paper, we have presented the main results of our extraction of a Flickr database. We've insisted on the heterogeneity of involvements, the diversity of users activities, the role of groups and social relations in the building of reputation and structuring the community. We want to conclude on the articulation between small and heavy users, which is one of the main features of large-scale social networking site. Even if the flexibility of Flickr platform brings together "Kodak" and "Snapr" Cultures, the main originality of Flickr is the way it facilitates conversations between amateurs of photography, who doesn't know each other in real life and who both play and gain reputation *with* photography. Our study shows that these users represent a small minority of Flickr registered accounts and nevertheless, they appear as a kind of leading group of the community. They create and animate new groups, comment other users' photos and tag with the collective purpose to create a specific space to share photos with others. A small minority of users, encouraging new activities (comments, groups discussion, tagging), has contributed to transform a photo storage space into an organized and living space of communication.

---

[13] in our two lists, these groups are **FLICKRGAYS** and **toycamera.com**.

[14] http://www.flickr.com/groups/scoreme/, /himom/, /scoring/ etc. for scores, /commentscommentscomments/, /comments/, /1on1/ etc. for… comments.

[15] There are actually also several concurrent groups intended for photos having been deleted in **DeleteMe!**.

We shall step further in future research by including the use profile and popularity of users while studying the various types of groups. Taking into account the role of the tags in the building of communities is also an important issue that was not investigated here.

## 6. ACKNOWLEDGMENTS

This work is part of a research project named *Autograph*, studying online auto-organized collectives and partly supported by the French ANR-Telecom national research program.

The authors thank Bertil Hatt and Maxime Crépel for many discussions on this work and Sébastien Bertrand for initially drawing their attention on Flickr. Of course nothing could have been done without Flickr's API and we hope this study will bring some new insights to the Flickr team.

## 7. REFERENCES

[1] Rheingold H**.**, *Virtual Communities*, Secker & Warburg, London, 1994**.**

[2] Aguiton C., Cardon D., The Strength of Weak Cooperation: an Attempt to Understand the Meaning of Web 2.0, *Communication & Strategies*, 65, 2007.

[3] Granovetter M. S., The Strength of Weak Ties, *The American Journal of Sociology*, vol 78, n° 6, may 1973.

[4] Cox A. M., Flickr: What is new in Web2.0?, in Proc. of Towards a social science of web2.0, Workshop "Towards as social science of Web2.0", University of York, 2007, http://www.shef.ac.uk/content/1/c6/04/77/66/flickr%20paper.pdf

[5] O'Reilly T., What is Web 2.0: design patterns and business models for the next generation of software, 2005 [online] Available at: http://www.oreillynet.com/pub/a/oreilly/tim/news/2005/09/30/what-is-web-20.html (August 26 2007).

[6] Fitzgerald M., « How We Did It: Stewart Butterfield and Caterina Fake, Co-founders, Flickr », December 2006. http://www.inc.com/magazine/20061201/hidi-butterfield-fake.html

[7] Van House N. A., Exhibition and Public Image-Sharing: Distant Closeness and Photo Exhibition, *CHI 2007*, San Jose, mai 2007.

[8] Cameron M., Naaman M., Boyd D., Davis M., HT06, Tagging Paper, Taxonomy, Flickr, Adacemic Article, ToRead, *Proceedings of the Seventeenth ACM Conference on Hypertext and Hypermedia*, ACM Press, Odense, Denmark, August 2006 : http://www.danah.org/papers/Hypertext2006.pdf

[9] Kumar R., Novak J., Tomkins A., Structure and Evolution of Online Social Networks, *KDD'06*, Philadelphia, August 20–23, 2006.

[10] Lerman K., Jones L. A., Social Browsing on Flickr, *ICWSM'2007*, Boulder, Colorado, 2007.

[11] Kirk D., Sellen A., Rother C., Wood K., Understanding Photowork, *Proceedings of the Conference on Human Factors in Computing Systems*, 2005. http://research.microsoft.com/sds/papers/Kirk_et_al_CHI06.pdf

[12] Miller A., Edwards K., Give and Take: A Study of Consumer Photo-Sharing Culture and Practice, *CHI 2007*, San Jose, may 2007. http://www.cc.gatech.edu/~keith/pubs/chi2007-photosharing.pdfhttp://www.oreillynet.com/pub/a/network/2005/02/04/sb_flckr.html

[13] Chalfen R. *Snapshot Versions of Life*, Bowling Green, Ohio, Bowling Green State University Popular Press, 1987.

[14] Okabe D., Ito M., Camera phones changing the definition of picture-worthy, *Japan Media review*, August 29, 2003.

[15] http://www.tomkinshome.com/papers/starpower/starpower.pdf

[16] Anthony D., Smith S., Williamson T., Explaining Quality in Internet Collective Goods: Zealots and Good Samaritans in the Case of Wikipedia, 2005. http://web.mit.edu/iandeseminar/Papers/Fall2005/anthony.pdf

[17] Mishne G., Glance N., Leave a Reply: An analysis of Weblog Comments, $3^{rd}$ *annual workshop on the Weblogging Ecosystem: aggregation, Analysis and Dynamics*, Edimburgh, WWW06 2006. http://www.blogpulse.com/www2006-workshop/papers/wwe2006-discovery-lin-final.pdf

[18] Lange Patricia G., Publicly private and Privately Public: Social Networking on YouTube, *Journal of Computer-Mediated Communication*, 13(1), article 18, 2007. http://jcmc.indiana.edu/vol13/issue1/lange.html

[19] Pissard N., Prieur P., Thematic vs. social networks in web 2.0 communities: A case study on Flickr groups, *Proc. of Algotel Conference*, 2007. http://hal.inria.fr/inria-00176954/en

[20] McDonald D. W., Visual Conversation Styles in Web Communities, *Proceedings of the $40^{th}$ Hawaii International Conference on System Sciences*, Kona, 2007. http://projects.ischool.washington.edu/mcdonald/papers/McDonald.HICSS-40.preprint.pdf